\documentclass[cernpreprint, atlasdraft=false, texlive=2017, UKenglish, texmf]{atlasdoc}
 
\usepackage[subcaption]{atlaspackage}
\usepackage{atlasbiblatex}
 
\usepackage{atlasphysics}
 
\graphicspath{{logos/}{figures/}}
 
\usepackage{ANA-STDM-2017-19-PAPER-defs}
\usepackage[utf8]{inputenc}
\usepackage{multirow}
\usepackage{hyperref}
 
\AtlasTitle{
Observation of electroweak production of two jets and a $Z$-boson pair
}

\AtlasAbstract{
Electroweak symmetry breaking explains the origin of the masses of elementary particles through their interactions with the Higgs field. Besides the measurements of the Higgs boson properties, the study of the scattering of massive vector bosons with spin one allows the nature of electroweak symmetry breaking to be probed. Among all processes related to vector-boson scattering, the electroweak production of two jets and a $Z$-boson pair is a rare and important one. Here we report the observation of this process from proton-proton collision data corresponding to an integrated luminosity of 139~\ifb{} recorded at a centre-of-mass energy of 13~\TeV{} with the ATLAS detector at the Large Hadron Collider. We consider two different final states originating from the decays of the $Z$-boson pair - one containing four charged leptons and the other containing two charged leptons and two neutrinos. The hypothesis of no electroweak production is rejected with a statistical significance of 5.7~$\sigma$, and the measured cross-section for electroweak production is consistent with the standard model prediction. In addition, we report cross-sections for inclusive production of a $Z$-boson pair and two jets for the two final states.
}

\AtlasRefCode{STDM-2017-19}
 
\PreprintIdNumber{CERN-EP-2020-016}

\AtlasJournal{Nature Physics}
\AtlasJournalRef{Nature Physics 19, 237–253 (2023)}
\AtlasDOI{https://doi.org/10.1038/s41567-022-01757-y}

\hypersetup{pdftitle={ATLAS document},pdfauthor={The ATLAS Collaboration}}
 
\begin{document}
 
\maketitle

\clearpage
 
\section{Probing electroweak symmetry breaking at the LHC}
\label{sec:intro}

Electroweak symmetry breaking (EWSB) plays a central role in the Standard Model (SM) of particle physics, as it explains the origin of elementary particle masses via the interactions of each particle with the Higgs field. Following the discovery of the Higgs boson~\cite{HIGG-2012-27,CMS-HIG-12-028}, the scrutiny of EWSB has become a primary focus of research at the Large Hadron Collider (LHC). In addition to direct measurements of the Higgs boson's properties, the scattering of two massive vector bosons (VBS) offers another key avenue to probe the EWSB mechanism~\cite{Lee:1977yc,Chanowitz:1985hj,Szleper:2014xxa}. As a result of the delicate configuration of quantum field interactions for the SM VBS processes~\cite{Lee:1977yc}, the presence of the Higgs boson is predicted to exactly cancel out the otherwise diverging VBS amplitudes at high energies and prevent unitarity violation at the \TeV\ scale. Any significant deviation from the predicted high-energy behaviour of VBS would point to new phenomena in the EWSB sector which are motivated by many plausible extensions to the SM~\cite{Contino:2011np,ArkaniHamed:2002qy,Djouadi:2013lra}. Moreover, VBS offers a sensitive means to search for anomalies in the weak-boson self-interactions~\cite{Eboli:2006wa,Bellan:2021dcy,Gomez-Ambrosio:2018pnl}, which are precisely predicted by the gauge theory in the SM.
 
The LHC provides an unprecedented opportunity to study the VBS process in proton--proton $(pp)$ collisions, due to the high collision energies and large luminosity. At the LHC, VBS occurs when two vector bosons ($V$) are radiated from the initial-state quarks in the colliding protons, and then scatter into another pair of vector bosons in the final state. The detector signature of VBS includes the decay products of the pair of outgoing bosons and a pair of hadronic jets ($j$), which originate from the deflection of the initial-state quarks that radiated the weak bosons. The most promising channel to measure VBS is the purely electroweak (EW) production of $VVjj$ (EW $VVjj$) in $pp$ collisions, in which the contributions from the non-VBS processes (such as triboson production) could be sufficiently suppressed with a proper choice of kinematic selections. Thus far, the EW $W^{\pm}W^{\pm}jj$ and $WZjj$ processes have been observed using LHC Run 2 data~\cite{Aaboud:2019nmv,Aaboud:2018ddq,Sirunyan:2017ret,CMS:2020gfh}, and no significant deviations from the SM predictions have been found. The Compact Muon Solenoid Collaboration has searched for EW $ZZjj$ production using 137~\ifb\ of 13~\TeV\ $pp$ collision data with an observed significance of 4.0 standard deviations~\cite{CMS-SMP-20-001}. Despite the small rate, EW $ZZjj$ production is of great interest, due to the low background and the unique feature of a fully reconstructed final state when both of the $Z$ bosons decay into charged leptons. The complete reconstruction of the final-state bosons provides maximal information in which the properties of the VBS process that are sensitive to EWSB can be probed. Furthermore, of all the measurements to date, VBS $ZZ$ production is uniquely sensitive to the possible anomalous interaction between four $Z$ bosons. This is forbidden at tree-level in the SM and the study of EW $ZZjj$ production is therefore a direct test of an important prediction of the electroweak theory. Finally, precision measurements of high-mass VBS $ZZ$ production also allow an almost model-independent measurement of the Higgs boson width. The Higgs width is precisely predicted by the SM and is sensitive to new phenomena in the Higgs sector. However, the current methods to extract the Higgs width (using the gluon--gluon fusion production mechanism) are known to fail for certain types of new phenomena and the use of the VBS production mechanism was proposed in order to alleviate this problem~\cite{Englert:2014aca}.
 
This article reports observation of EW $ZZjj$ production at the LHC, as well as a measurement of the cross-sections of the inclusive (EW and non-EW) $ZZjj$ processes. The set of 13~\TeV\ $pp$ collision data recorded by the ATLAS experiment during the LHC Run 2 is used. The search is performed in two final states where both $Z$ bosons decay leptonically: in final states with either four charged leptons and two jets (\lllljj), or two charged leptons, two neutrinos and two jets (\llvvjj). The definition of the signal region is optimised to suppress the reducible backgrounds coming from processes with different final states. Multivariate discriminants (MDs) are used to further separate the EW signal from the remaining backgrounds, including both the reducible ones and the irreducible non-EW $ZZjj$ process, which contains two strong interactions at the lowest order in perturbation theory and is referred to as quantum chromodynamics (QCD) $VVjj$ production. Figure~\ref{fig:diagram} depicts the typical diagrams for both the EW VBS and QCD $ZZjj$ processes. These MDs exploit the characteristics of VBS production, such as a large separation in rapidity between the two jets (\dyjj) as well as a large invariant mass of the jet pair (\mjj). The production of $ZZjj$ in which one or both $Z$ bosons decay into electrons or muons via $\tau$-leptons is considered as signal, but it makes a negligible contribution to the selected event sample.

\begin{figure}[!htbp]
\begin{center}
\begin{subfigure}[b]{0.22\textwidth}
\includegraphics[width=\textwidth]{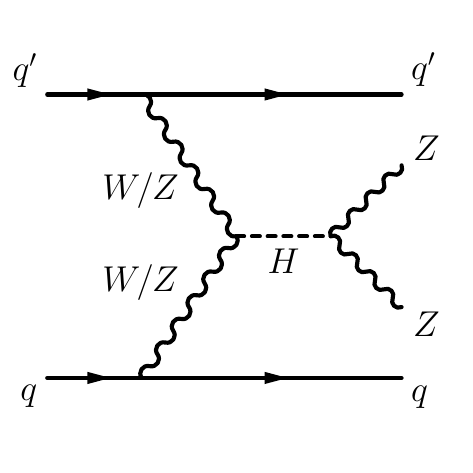}
\caption{}
\end{subfigure}
\begin{subfigure}[b]{0.22\textwidth}
\includegraphics[width=\textwidth]{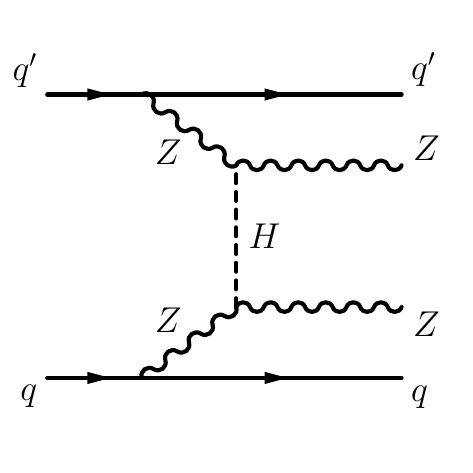}
\caption{}
\end{subfigure}
\begin{subfigure}[b]{0.22\textwidth}
\includegraphics[width=\textwidth]{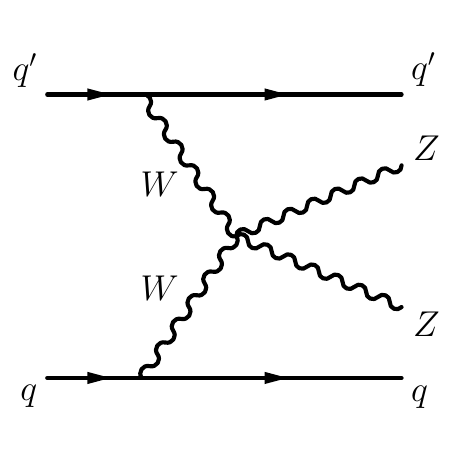}
\caption{}
\end{subfigure}
\begin{subfigure}[b]{0.22\textwidth}
\includegraphics[width=\textwidth]{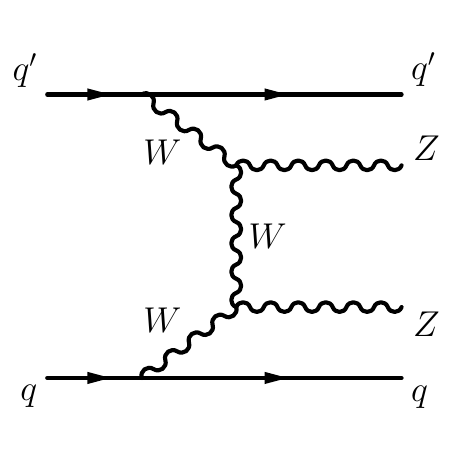}
\caption{}
\end{subfigure}
\\
\begin{subfigure}[b]{0.19\textwidth}
\includegraphics[width=\textwidth]{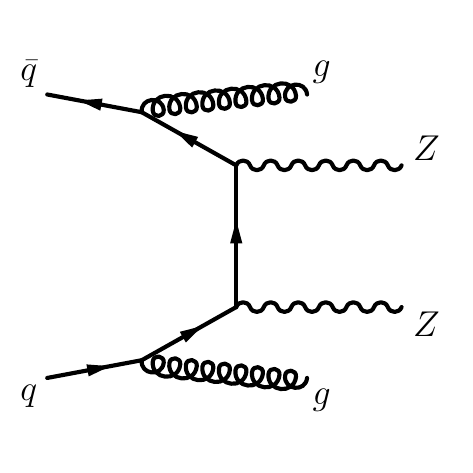}
\caption{}
\end{subfigure}
\begin{subfigure}[b]{0.19\textwidth}
\includegraphics[width=\textwidth]{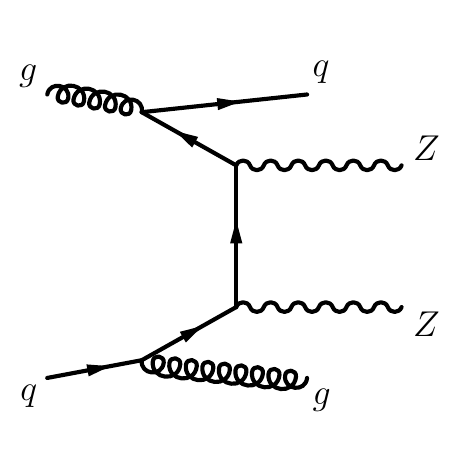}
\caption{}
\end{subfigure}
\begin{subfigure}[b]{0.19\textwidth}
\includegraphics[width=\textwidth]{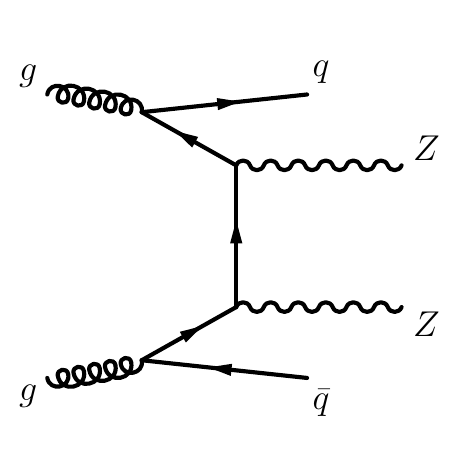}
\caption{}
\end{subfigure}
\begin{subfigure}[b]{0.19\textwidth}
\includegraphics[width=\textwidth]{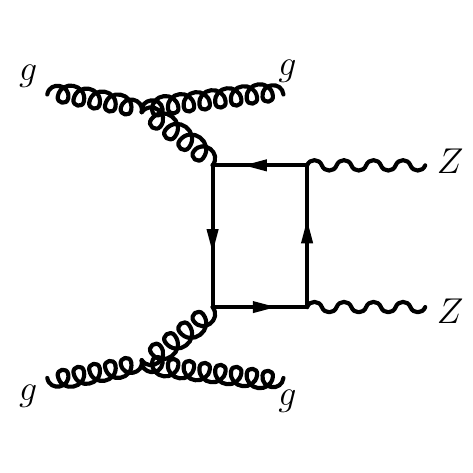}
\caption{}
\end{subfigure}
\begin{subfigure}[b]{0.19\textwidth}
\includegraphics[width=\textwidth]{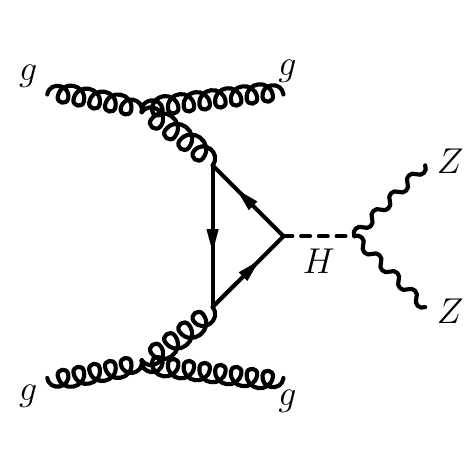}
\caption{}
\end{subfigure}
\\
\end{center}
\caption{\textbf{Typical Feynman diagrams for the production of $ZZjj$.}
a-d. The relevant EW VBS diagrams are shown in the first row for (a) the s-channel and (b) the t-channel production through a Higgs boson, (c) the weak-boson self-interaction process, and (d) the production through exchange of a $W$ boson.
The relevant QCD diagrams are shown in the second row for (e to g) the tree-level production with different quark and gluon initial states, (h) the box diagram without a Higgs boson, and the (i) the triangle diagram through a Higgs boson.
In these diagrams, straight lines represent quarks ($q$, $q'$), spiral lines represent gluon ($g$), wave lines represent the $W$ and $Z$ bosons, and the dashed lines represent the Higgs boson.
}
\label{fig:diagram}
\end{figure}
 

\section{Experimental apparatus}
\label{sec:detector}

The ATLAS experiment~\cite{PERF-2007-01,ATLAS-TDR-2010-19,Abbott:2018ikt} at the LHC uses a multipurpose particle detector with a forward--backward symmetric cylindrical geometry and a near \(4\pi\) coverage in solid angle. ATLAS uses a right-handed coordinate system with its origin at the nominal interaction point in the centre of the detector and the \(z\)-axis along the beam pipe. The \(x\)-axis points from the interaction point to the centre of the LHC ring, and the \(y\)-axis points upwards. Cylindrical coordinates \((r,\phi)\) are used in the transverse plane, \(\phi\) being the azimuthal angle around the \(z\)-axis. The pseudorapidity is defined in terms of the polar angle \(\theta\) as \(\eta = -\ln \tan(\theta/2)\). The angular distance between two physics objects is measured in units of \(\Delta R \equiv \sqrt{(\Delta\eta)^{2} + (\Delta\phi)^{2}}\). The ATLAS detector consists of an inner tracking detector surrounded by a thin superconducting solenoid providing a \SI{2}{\tesla} axial magnetic field, electromagnetic and hadron calorimeters, and a muon spectrometer. The inner tracking detector covers the pseudorapidity range \(|\eta| < 2.5\). It consists of silicon pixel, silicon microstrip, and transition radiation tracking detectors. Lead/liquid-argon (LAr) sampling calorimeters provide electromagnetic energy measurements with high granularity. A steel/scintillator-tile hadron calorimeter covers the central pseudorapidity range (\(|\eta| < 1.7\)). The endcap and forward regions are instrumented with LAr calorimeters for electromagnetic and hadronic energy measurements up to \(|\eta| = 4.9\). The muon spectrometer covers the pseudorapidity range \(|\eta| < 2.7\) and is based on three large air-core toroidal superconducting magnets with eight coils each. The field integral of the toroids ranges between \num{2.0} and \SI{6.0}{Tm} across most of the detector. The muon spectrometer includes a system of precision tracking chambers and fast detectors for triggering. A two-level trigger system~\cite{TRIG-2016-01} is used to select events for offline analysis. The first-level trigger is implemented in hardware and uses a subset of the detector information. This is followed by the software-based high-level trigger, which reduces the event selection rate to about 1~kHz.

\section{Data and simulation}
\label{sec:datamc}

The data for this analysis were recorded using single-lepton and multi-lepton triggers, corresponding to an integrated luminosity of 139~\ifb. The overall trigger efficiency for the inclusive $ZZjj$ events selected for this analysis ranges from 95\% to 99\%{} for the inclusive sample of all final states considered.
 
The EW $ZZjj$ production was modelled using the \textsc{Powheg-Box}~v2~event generator~\cite{PowhegV2} with matrix elements (ME) calculated at the next to leading order (NLO) in perturbative QCD (pQCD) and with the NNPDF3.0LO~\cite{ball2015parton} parton distribution functions (PDF). The contributions from triboson and VH processes in \lllljj and \llvvjj channels were estimated using the \MGNLO~2.6.1 event generator~\cite{Alwall:2014hca} with ME calculated at leading order (LO) in pQCD with the NNPDF3.0LO PDF. Reweighting factors were calculated as a function of \mjj from the \MGNLO events and applied to the POWHEG-V2 events. The effect is found to be below a few percent level. The QCD $ZZjj$ production was modelled using \textsc{Sherpa}~2.2.2~\cite{Gleisberg:2008ta} with the NNPDF3.0NNLO~\cite{ball2015parton} PDF. The events with up to one outgoing parton were generated at NLO in pQCD, while those with two or three partons were modelled with LO accuracy. The production of $ZZjj$ from the gluon--gluon initial state with a four-fermion loop or with the exchange of a Higgs boson was generated separately. This process, referred to as the $ggZZjj$ process, was modelled using \textsc{Sherpa}~2.2.2 with the NNPDF3.0NNLO PDF and \textsc{gg2VV}~\cite{Kauer:2013qba} with the CT10NNLO~\cite{Gao:2013xoa} PDF in the \lllljj and \llvvjj channels, respectively, and normalised to a calculation accurate to NLO in pQCD. The leptonic decays of $Z$ bosons are included in the simulation. Interference between EW and QCD $ZZjj$ was modelled with \MGNLO~2.6.1 calculated at LO and is treated as systematic on the predicted EW process. The effect is far smaller than the statistical uncertainty from data.
 
The production of $WWjj$ and $WZjj$ with the subsequent leptonic decays of vector bosons were modelled with \textsc{Sherpa}~2.2.2. Diboson processes with the subsequent semileptonic decays were modelled using \textsc{Powheg-Box}~v2~\cite{Frixione:2007nw}. Triboson production not in the \lllljj nor \llvvjj channels was modelled using \textsc{Sherpa}~2.2.2. For top-quark pair production, \textsc{Powheg-Box}~v2 was used. The production of single top quarks was simulated using \textsc{Powheg-Box}~v1~\cite{Alioli:2009je,Frederix:2012dh,Re:2010bp}. The production of ttbar~in association with vector bosons ($ttV$) was modelled with \MGNLO~2.3.3 for $ttW$, with \textsc{Sherpa}~2.2.1 and \MGNLO~2.3.3 for $ttZ$, and with \MGNLO~2.2.2 for $ttWW$. The \Zjet processes were modelled using \textsc{Sherpa}~2.2.1.
 
The above theoretical calculations are accurate to a given order in perturbation theory for partonic final states. To correctly model the hadronic final state that interacts with the detector, parton showering, hadronisation and underlying-event algorithms were applied to the partonic final states predicted from each calculation. Those were modelled with \textsc{Pythia}~8.186~\cite{Sjostrand:2007gs} using the NNPDF2.3LO~\cite{Ball:2012cx} PDF and the A14 set of tuned parameters~\cite{ATL-PHYS-PUB-2014-021} for all the samples except for the ones from \textsc{Sherpa}, where those were simulated within the \textsc{Sherpa} program.
 
All samples were passed through a detailed simulation of the ATLAS detector~\cite{SOFT-2010-01} based on \textsc{Geant4}~\cite{Agostinelli:2002hh}, to produce predictions that can be directly compared with the data. Furthermore, simulated inelastic $pp$ collisions were overlaid to model additional $pp$ collisions in the same and neighbouring bunch crossings (pile-up)~\cite{ATLAS:2021ija}. Simulated events were reweighted to match the pile-up conditions in the data. All simulated events were processed using the same reconstruction algorithms as used in data.

\section{Event selection}
\label{sec:sel}

The selection of the \lllljj and \llvvjj events relies on multiple physics objects, including electrons, muons, and jets. The signal region (SR) is defined with a set of selection criteria which were optimised to preferentially select the EW $ZZjj$ events.
 
Events are first required to have a collision vertex associated with at least two tracks each with transverse momentum (\pt) $>0.5$~\GeV. The vertex with the highest sum of $p_{\mathrm{T}}^{2}$ of the associated tracks is referred to as the primary vertex.
 
Muons are identified by tracks reconstructed in the muon spectrometer and are matched to tracks reconstructed in the inner detector (ID). In the region $2.5<|\eta|<2.7$, muons can also be identified by tracks from the muon spectrometer alone, and these are called stand-alone muons. Identified muons are required to have $\pT > 7$~\GeV{}. In the gap region ($|\eta|<0.1$) in muon spectrometer, muons are identified by an track from inner tracking detector with $\pT > 15$~\GeV{} associated with a compatible calorimeter energy deposit, and are called calorimeter-tagged muons. Muons are required to have $|\eta| < 2.7$ and satisfy the `loose' identification criterion~\cite{PERF-2015-10} in the \lllljj~channel, while they must satisfy $|\eta| < 2.5$ and the `medium' identification in the \llvvjj~channel. Electrons are reconstructed from energy deposits in the electromagnetic calorimeter matched to a track in the ID. Candidate electrons must have $\pT > 7$~\GeV{} and $|\eta| < 2.47$, and satisfy the `loose' and `medium' identification criteria~\cite{PERF-2017-01} in the \lllljj~and \llvvjj~channels, respectively. All electrons and muons must be isolated and satisfy the `FixedCutLoose' and `loose' isolation criteria~\cite{PERF-2015-10,PERF-2017-01} in the \lllljj and \llvvjj channels, respectively. Furthermore, electrons (muons) are required to have associated tracks satisfying $|d_0/\sigma_{d_0}|<5~(3)$ and $|z_0\times\sin\theta|<0.5$~mm, where $d_0$ is the transverse impact parameter relative to the beam line, $\sigma_{d_0}$ is its uncertainty, and $z_0$ is the longitudinal impact parameter relative to the primary vertex.
 
Jets are reconstructed from clusters of calorimeter energy deposits using the anti-$k_t$ algorithm~\cite{antikt_algorithm,Fastjet} with radius parameter $R = 0.4$. The jet energy scale is calibrated using simulation and further corrected with in situ methods~\cite{Aaboud:2017jcu}. A jet-vertex tagger~\cite{PERF-2014-03} is applied to jets with $\pT<60$~\GeV\ and $|\eta|<2.4$ to preferentially suppress jets that originated from pile-up. In addition, jets containing $b$-hadrons ($b$-jets) are identified using a multivariate $b$-tagging algorithm~\cite{Aad:2019aic}. The chosen $b$-tagging algorithm has an efficiency of 85\%{} for $b$-jets and a rejection factor of 33 against light-flavour jets.
 
An overlap-removal procedure detailed in Ref.~\cite{Aad:2016eki} is applied to the selected leptons and jets in the \llvvjj channel, to avoid ambiguities in the event selection and in the energy measurement of the physics objects. A similar approach is adopted in the \lllljj channel, except that leptons are given a higher priority to be kept when overlapping with jets, to enhance the selection efficiency.
 
The neutrinos in the \llvvjj final state do not interact with the ATLAS detector and cannot be reconstructed. Their presence is identified using the missing transverse momentum vector ($\vec{E}_{\mathrm{T}}^{\mathrm{miss}}$), which is computed as the negative of the vector sum of transverse momenta of all the leptons and jets, as well as the tracks originating from the primary vertex but not associated with any of the leptons or jets~\cite{ATL-PHYS-PUB-2015-027}. The statistical significance of \met (\met-significance) is calculated using resolution information of physics objects used in the \met~reconstruction~\cite{ATLAS-CONF-2018-038}.
 
In the \lllljj channel, quadruplets of leptons are formed by selecting two opposite-sign, same-flavour lepton pairs, where the leptons are required to be separated from each other by $\Delta R >0.2$. At most one muon is allowed to be a stand-alone or calorimeter-tagged muon, and the three leading leptons must have $\pt > 20$, 20 and 10~\GeV{}, respectively. All the $\ell^+\ell^-$ pairs are required to have an invariant mass ($m_{\ell^+\ell^-}$) greater than 10~\GeV{}, to reject events from low-mass resonances. If multiple quadruplets are found, the one that minimises the sum of the differences between the dilepton masses and the nominal $Z$-boson mass, $|m_{\ell^+\ell^-} - m_Z| + |m_{\ell^{'+}\ell^{'-}} - m_Z|$, is selected. The dilepton masses are required to be within the range 66--116~\GeV{}.
 
In the \llvvjj channel, candidate events are required to have one opposite-sign, same-flavour lepton pair with $m_{\ell^+\ell^-}$ in the range from 80 to 100~\GeV{}, and the leading (sub-leading) lepton must have $\pt > 30~(20)$~\GeV{}. Events with $b$-tagged jets or additional leptons ($\pt > 7$~\GeV{} and satisfying the `loose' requirement) are rejected to reduce the background contributions from \ttbar~and $WZ$ events. Events are required to have an \met-significance greater than 12 to suppress the background from \Zjet processes.
 
In both channels, the two jets with the highest \pt and satisfying a negative product of jet rapidities ($y_{j_1} \times y_{j_2})$ are selected. In the \lllljj channel the jets are required to have $\pT > 30$~(40)~\GeV{} in the $|\eta| < 2.4$ $(2.4 < |\eta| < 4.5)$ region, while in the \llvvjj channel the leading (sub-leading) selected jet is required to have $\pT > 60$ (40)~\GeV{}. Finally, to further suppress background contributions, \dyjj is required to be greater than two, and \mjj is required to be greater than 300~\GeV{} and 400~\GeV{} in the \lllljj~and \llvvjj~channels, respectively. The harsher jet requirement in the \llvvjj channel is optimised to suppress the larger contamination from reducible backgrounds.
 
After selection, the resulting observed and expected yields are listed in Table~\ref{tab:yield_prefit}, where in total 127 and 82 data events are selected in the \lllljj~and \llvvjj~channels, respectively. Several control regions (CRs), defined with dedicated selections optimised to enhance the fractions of background events, are defined to constrain the contributions from the various background processes. The kinematic distributions from both channels, including the \mjj spectra in the \lllljj SR, QCD $ZZjj$ CR and \llvvjj SR, as well as the invariant mass of the four-lepton system ($m_{ZZ}$) in the \lllljj SR, are presented in Figure~\ref{fig:kine}. The background estimates, dedicated CRs and various sources of experimental and theoretical uncertainties are discussed in the following two sections.
 
The number of events in data is found to be consistent with the SM prediction including the EW $ZZjj$ contribution.
 
\begin{table}[!htbp]
\caption{
\textbf{Observed data and expected event yields in 139~\ifb{} of data in the \lllljj and \llvvjj signal regions.}
All the minor backgrounds are summed together as `Others', and the $WWjj$ and \ttbar~processes are referred to as the \nonresll~backgrounds.
Uncertainties in the predictions include both the statistical and systematic components on the predicted yields before fit.
}
\sisetup{
table-number-alignment = center,
table-align-uncertainty=true
}
\begin{center}
\begin{tabular}{
l
S[table-format = 3.1]@{$\,\pm\,$}
S[table-format = 2.1]
S[table-format = 2.1]@{$\,\pm\,$}
S[table-format = 1.2]
}
\hline
Process                 & \multicolumn{2}{c}{\lllljj}    & \multicolumn{2}{c}{\llvvjj}           \\
\hline
EW $ZZjj$               &  31.4 &  3.5 & 15.0 & 0.8  \\
QCD $ZZjj$              &  77   & 25   & 17.2  & 3.5  \\
QCD $ggZZjj$            &  13.1 &  4.4 &  3.5  & 1.1  \\
\hline
Non-resonant-$\ell\ell$ &  \multicolumn{2}{c}{--} & 21.4 & 4.8  \\
$WZ$                    &  \multicolumn{2}{c}{--} & 24.6 & 1.1  \\
Others                  &   3.2 &  2.1 & 1.2  & 0.9  \\
\hline
Total                   & 124   & 26   & 82.9 & 6.4  \\
\hline
Data                    &  \multicolumn{2}{l}{127}    & \multicolumn{2}{l}{82}  \\
\hline
\end{tabular}
\end{center}
\label{tab:yield_prefit}
\end{table}
 
\begin{figure}[!htbp]
\begin{center}
 
\begin{subfigure}[b]{0.45\textwidth}
\centering
\includegraphics[width=\textwidth]{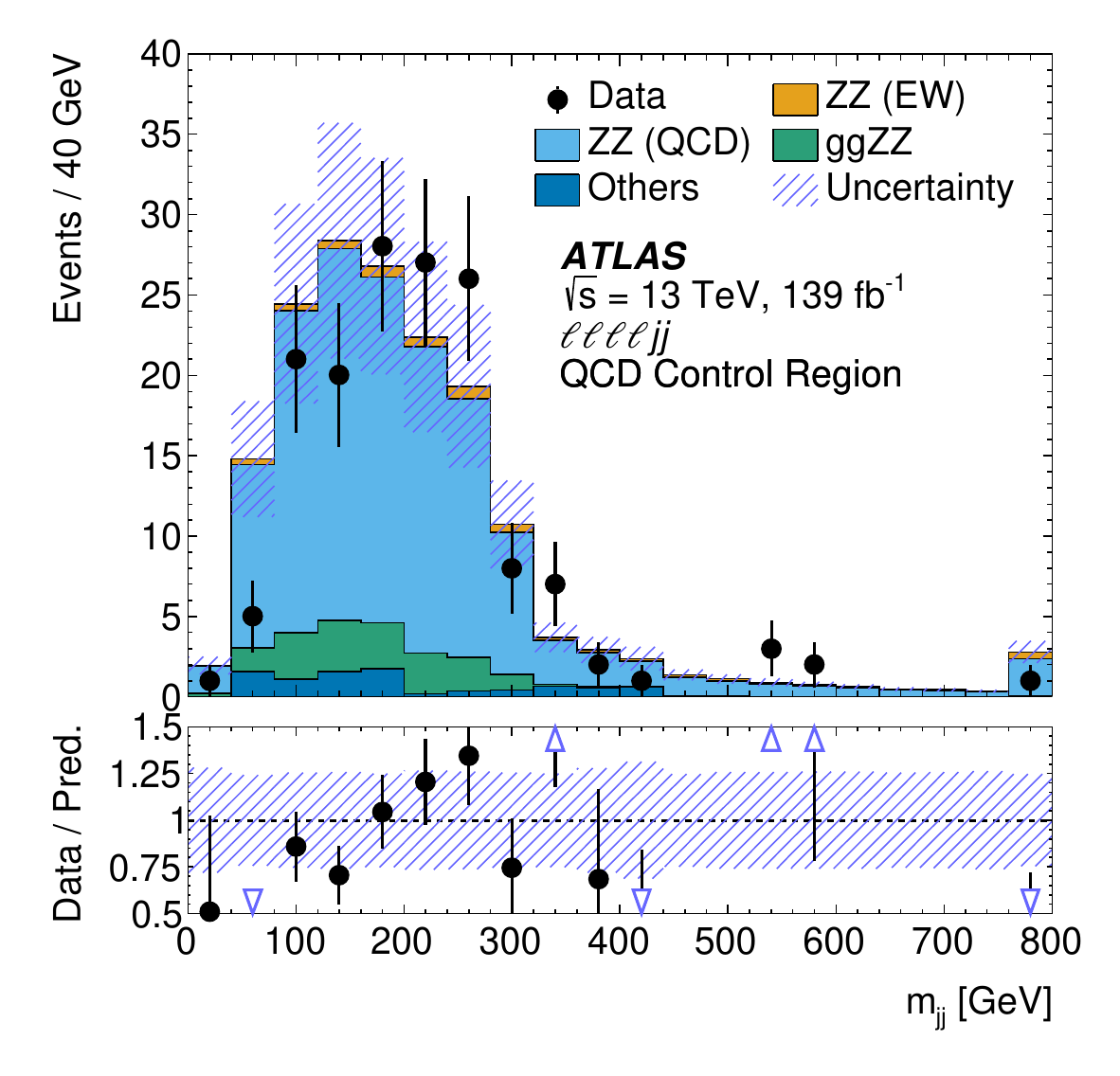}
\caption{}\label{fig:kine_a}
\end{subfigure}
\begin{subfigure}[b]{0.45\textwidth}
\centering
\includegraphics[width=\textwidth]{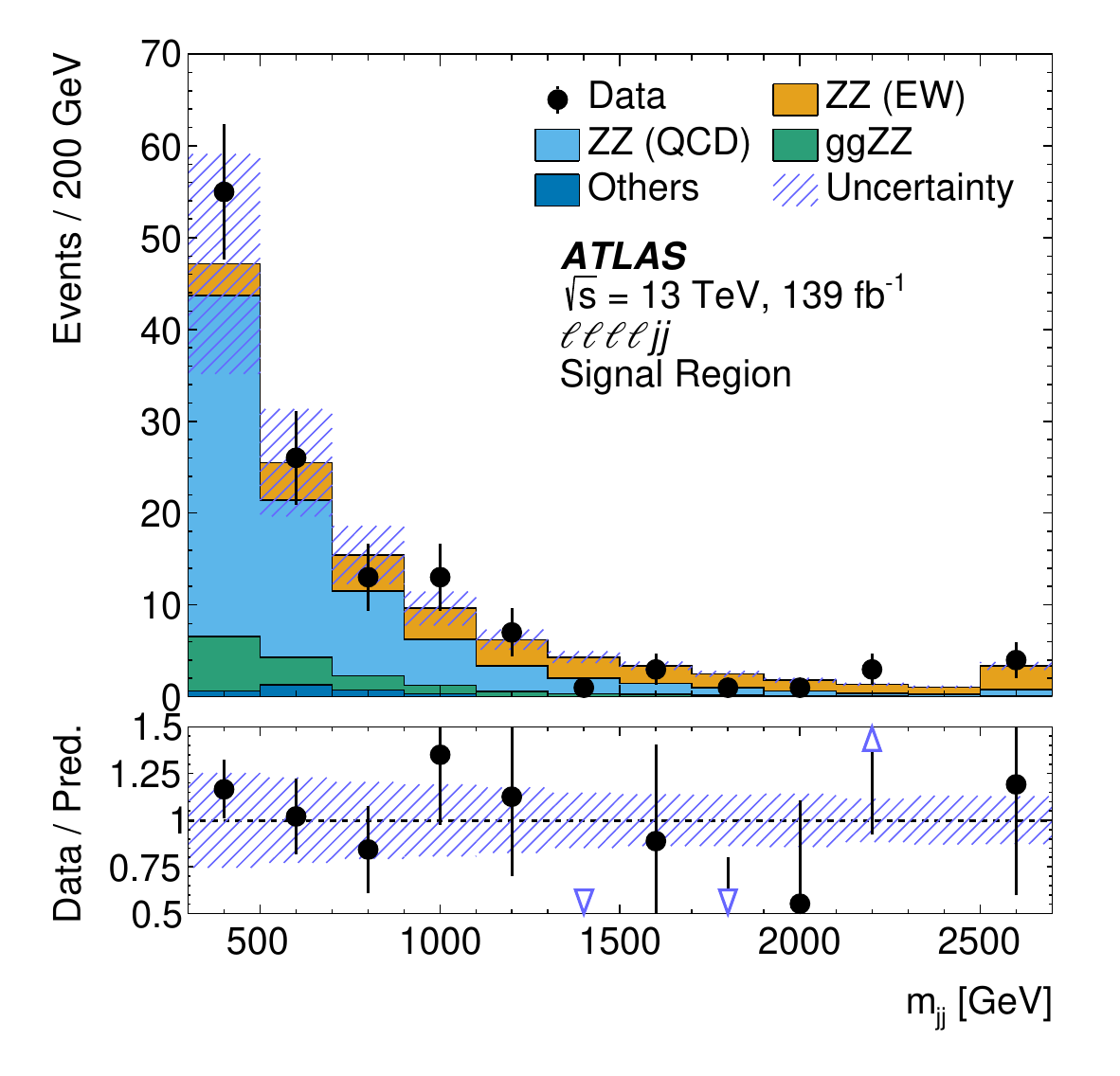}
\caption{}\label{fig:kine_b}
\end{subfigure}
 
\begin{subfigure}[b]{0.45\textwidth}
\centering
\includegraphics[width=\textwidth]{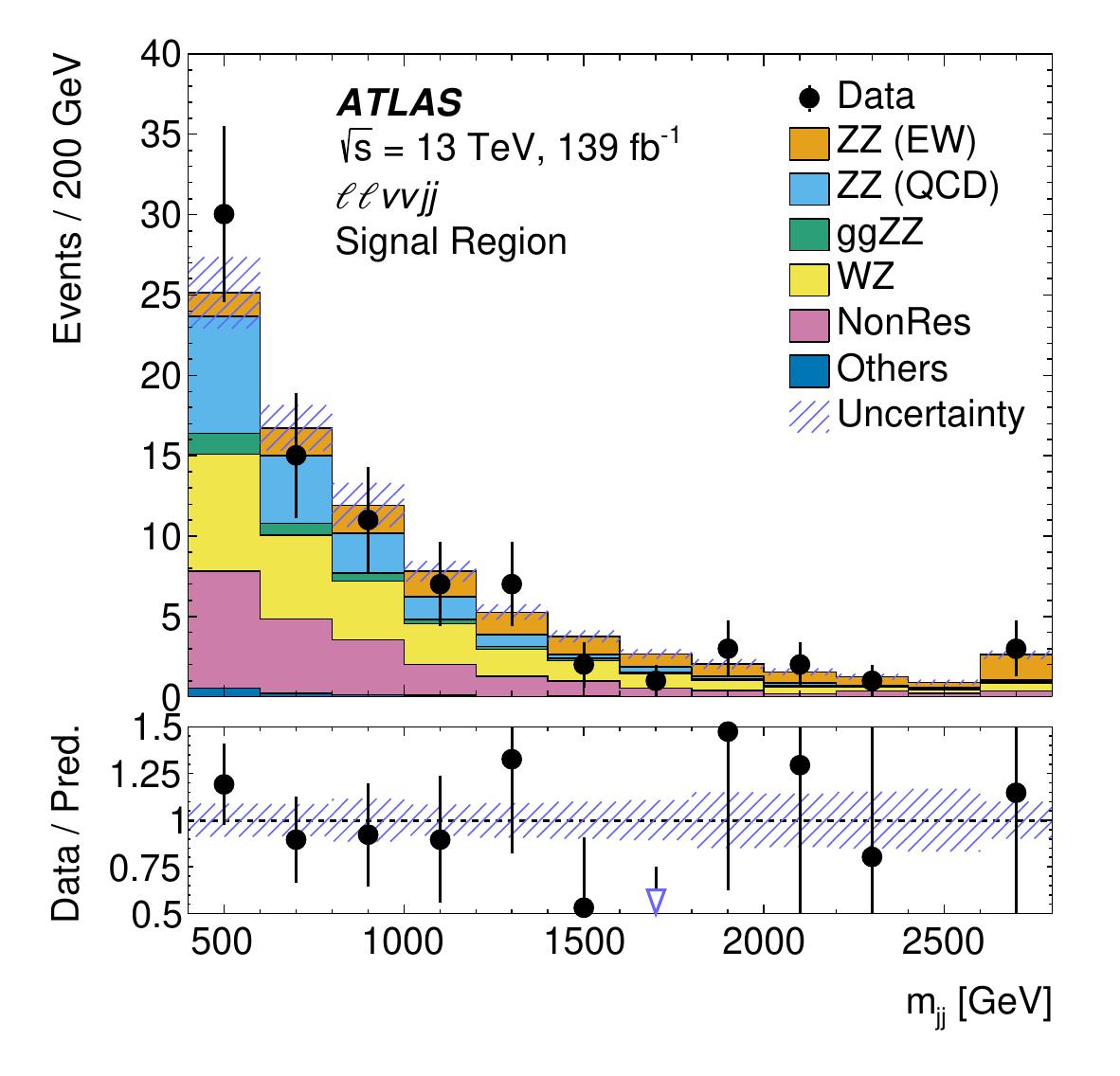}
\caption{}\label{fig:kine_c}
\end{subfigure}
\begin{subfigure}[b]{0.45\textwidth}
\centering
\includegraphics[width=\textwidth]{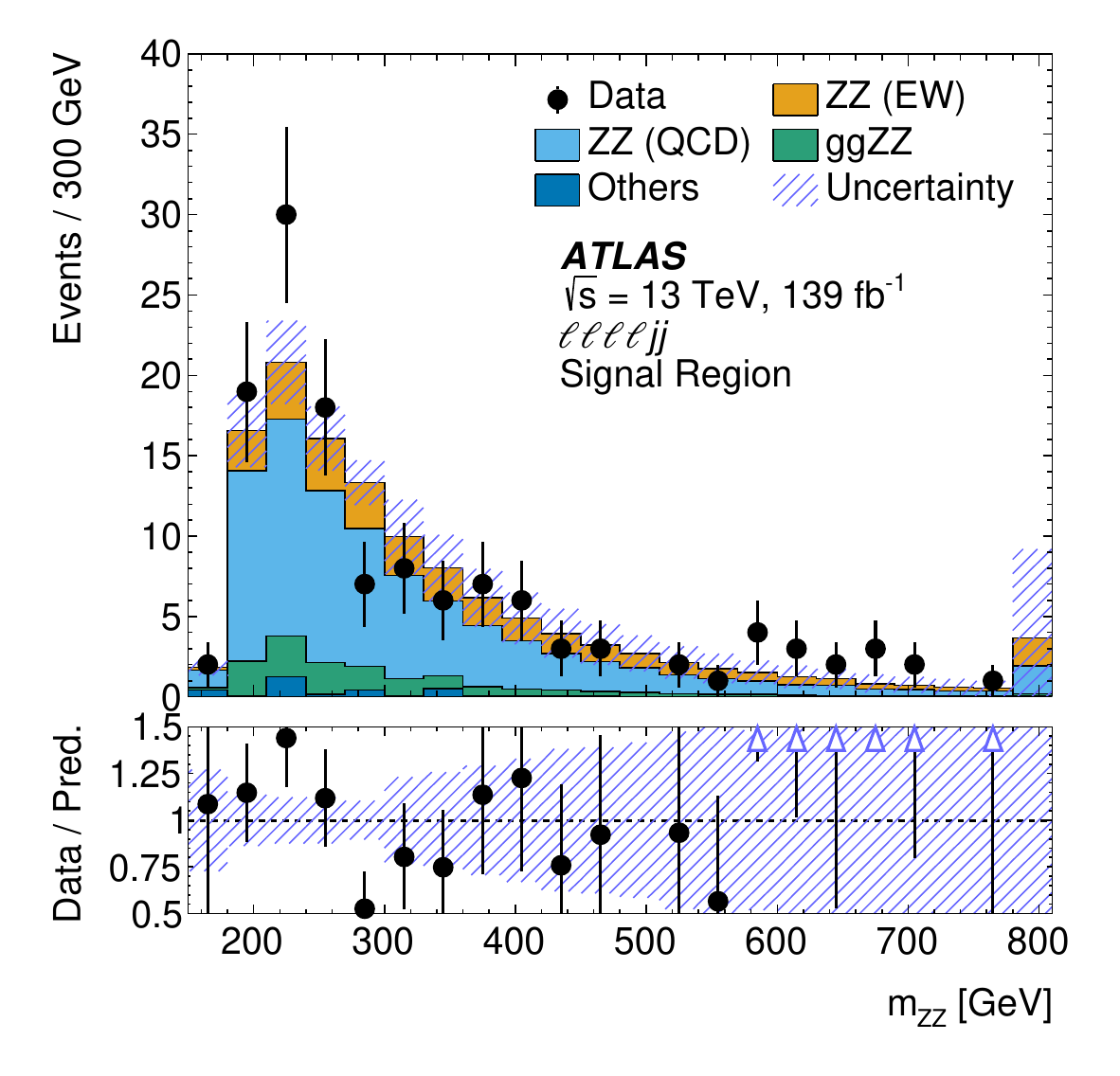}
\caption{}\label{fig:kine_d}
\end{subfigure}
 
\end{center}
\caption{\textbf{Observed and expected distributions.} a-d. The \mjj distributions in the (a) \lllljj QCD CR and
the (b) \lllljj and (c) \llvvjj signal regions,
as well as the $m_{ZZ}$ distribution in the (d) \lllljj signal region.
The error bands represent standard deviations and include the expected experimental and theoretical systematic uncertainties.
The contributions from the QCD and EW production of $ZZjj$ events are scaled by 0.99 and 0.92, respectively,
which correspond to the observed normalisation factors in the statistical fit to the combined channel (Table~\ref{tab:fit_result}).
The ZZ (EW), ZZ (QCD) and ggZZ represent contributions from EW, non-$gg$ QCD and $gg$ QCD $ZZjj$ processes, respectively.
The WZ represents contribution from $WZjj$ process.
All the minor backgrounds are summed together as `Others',
and the $WWjj$ and \ttbar~processes are referred to as `NonRes'.
The last bin includes the overflow events.
The statistical uncertainties of the data are shown as error bars.
The open arrows represent the out-of-range markers.
The horizontal bin width is indicated on the vertical axis legend.
}
\label{fig:kine}
\end{figure}
 

\section{Background estimation}
\label{sec:bkg}

The backgrounds arise from two kinds of processes, one with final-state particles that are the same as those in the signal process, and the other one where one or more lepton candidates are misidentified in data.
 
In the \lllljj channel, the largest background arises from the QCD $ZZjj$ process, which has an identical final state to the EW $ZZjj$ process. The kinematic properties of the QCD $ZZjj$ background are estimated using the simulated events described in Section~\ref{sec:datamc}. However, the simulation is normalised to data in a dedicated EW-suppressed control region, defined by reversing either the \mjj or the \dyjj selection criteria. Furthermore, the modelling of the kinematic properties of the QCD $ZZjj$ simulation is validated in an additional EW-suppressed validation region, which is defined by requiring the centrality to be larger than 0.5 for at least one of the selected $Z$ bosons. The centrality is a variable that estimates the position of a $Z$-boson with respect to the rapidity span of the two outgoing hadronic jets~\cite{STDM-2017-26}. The EW $ZZjj$ contribution is less than 4\% in the additional EW-suppressed validation region. More than 70\% of the QCD $ZZjj$ events in the \mjj $>$ 300~\GeV{} region is not overlapping with the events in the additional EW-suppressed validation region. Good agreement between data and simulation is found for most kinematic distributions. The impact on the signal extraction of a potential mismodelling of the \mjj distribution in the QCD $ZZjj$ simulation (seen in previous analyses~\cite{STDM-2013-02,STDM-2014-11,STDM-2016-09}) is explicitly tested, by reweighting the QCD $ZZjj$ simulation in the SR using an \mjj-dependent correction factor that is defined as the ratio of data to simulation in the EW-suppressed validation region at high centrality. The signal extracted using the reweighted and nominal QCD $ZZjj$ simulations are found to be in agreement when considering the statistical uncertainty on the \mjj-dependent correction itself.
 
Small background contributions from \Zjet, top-quark and $WZjj$ processes may contain misidentified leptons and are estimated using a method similar to that described in Ref.~\cite{Aaboud:2017rwm}, where the lepton misidentification is measured in data regions with enhanced contributions from \Zjet and top-quark processes. Minor background contributions from triboson and $ttV$ production are estimated from simulation. All of those backgrounds collectively yield an estimated contribution of about 3\% to the selected data sample in the \lllljj channel.
 
In the \llvvjj channel, the normalisation and kinematic properties of QCD $ZZjj$ processes are modelled from simulation due to a large contamination from other processes, which are considered in dedicated CRs. The $WZjj$ background, with one lepton produced outside of the detector acceptance, is estimated using a data CR defined by requiring three selected leptons and a looser event selection, following the methodology explained in Ref.~\cite{Aaboud:2017bja}. The simulation is found to overestimate the $WZjj$ contribution by 23\% in this CR, and therefore the $WZjj$ yield in the SR is scaled by 0.81. The $WZjj$ distribution of the MD in the SR is evaluated from simulation, with the contributions from EW $WZjj$ processes scaled by 1.77, corresponding to the difference between data and simulation observed in a previous analysis, in a similar phase space~\cite{Aaboud:2018ddq}, where the overall normalisation factor is found to be consistent with the one derived in this article. The $WWjj$ and \ttbar~processes are referred to as the \nonresll~backgrounds, since they contain a lepton pair not originating from a $Z$ or $\gamma^{*}$ boson. The \nonresll background is estimated using a CR defined in data by applying the same selection as in the SR with the exception that an $e\mu$ pair is required, following the methodology explained in Ref.~\cite{Aaboud:2017bja}. The MD distribution in the SR for the \nonresll~process is estimated from simulation.
 
The \Zjet background is largely suppressed, and the yield is evaluated by extrapolating the low \met-significance region distribution in data to the high \met-significance region using an exponential function, while the MD distribution in the SR is modelled by simulation. Uncertainties are assigned to account for variations in the fitting functions as well as differences between estimated and simulated yields and distributions. Small background contributions from triboson and $ttV$ production are modelled with simulation. All of those backgrounds collectively yield an estimated contribution of about 1\% to the selected data sample in the \llvvjj channel.

\section{Experimental and theoretical uncertainties}
\label{sec:unc}

Systematic uncertainties associated with the prediction of each signal and background process are estimated. These uncertainties are either experimental or theoretical in nature, due to imperfect modelling of the detector in the simulation or the underlying physics of each process.
 
The major experimental uncertainties originate from the luminosity uncertainty, the energy measurements of leptons and jets, and the lepton reconstruction and selection efficiencies. Smaller experimental uncertainties are also considered, such as those due to the trigger selection efficiency, the calibration of the \met~soft-term, the pile-up correction, and the $b$-jet identification efficiency. Overall, the total experimental uncertainty in the predicted yields is about 10\% and 5\% in the \lllljj~and \llvvjj~ channels, respectively. The dominant uncertainties originate from jet and lepton calibration. The uncertainty in the combined 2015--2018 integrated luminosity is 1.7\%~\cite{ATLAS-CONF-2019-021}, obtained using the LUminosity Cherenkov Integrating Detector-2 detector \cite{LUCID2} for the primary luminosity measurements.
 
In addition, the uncertainties in the predicted yields from non-$ZZjj$ backgrounds are dominated by the statistical uncertainties from data in the dedicated CRs, which are about 15\% for the overall backgrounds. The $WZjj$ shape uncertainty originates from experimental and theoretical uncertainties as well as from the uncertainty in the quoted EW $WZjj$ cross-section measurement. The \nonresll shape uncertainty is estimated by comparing the MD distributions from data and simulation. In addition, an uncertainty is assigned to the QCD $ZZjj$ processes by comparing the MD distributions in low and high pile-up conditions, to account for a potential mismodelling of pile-up in simulation. This uncertainty is only considered for the QCD $ZZjj$ background, given its dominant impact in this analysis and the non-negligible probability for QCD $ZZjj$ events to contain a pile-up jet. The differences in the predicted yields in different MD regions are below 10\% except in the last bin in the QCD $ZZjj$ CR where it reaches 50\% due to the statistical uncertainty of the simulated events.
 
The theoretical uncertainties of the EW and QCD $ZZjj$ processes include the uncertainties from PDFs, QCD scales, strong coupling constant ($\alpha_\textrm{S}$), parton showering and hadronisation. Those are estimated with the \MGNLO~2.6.1 for EW $ZZjj$ and \textsc{Sherpa}~2.2.2 for QCD $ZZjj$ processes. The PDF uncertainty is estimated following the PDF4LHC~\cite{Butterworth:2015oua} procedure. The effect of the QCD scale uncertainty is estimated by varying the renormalisation and factorisation scales following the procedure described in Ref.~\cite{Aaboud:2019lgy}. The parton showering and hadronisation uncertainty is estimated by comparing the nominal \textsc{Pythia}~8 parton showering with the alternative \textsc{Herwig}~7~\cite{Bellm:2015jjp, Bahr:2008pv} algorithm. The effect of the $\alpha_\textrm{S}$ uncertainty is estimated by varying the $\alpha_\textrm{S}$ value by $\pm0.001$. The total theoretical uncertainties in the reconstructed event yields for the EW and QCD $ZZjj$ process are estimated to be about 10\% and 30\%, respectively. Those uncertainties have been checked to confirm the nuisance parameters associated to them are not over-constrained with current dataset, therefore, it is inferred that this analysis is not sensitive to theoretical uncertainties beyond the LO. The interference effect between the EW and QCD processes is studied using \MGNLO~2.6.1 interfaced to \textsc{Pythia}~8.186 and found to make a relative contribution (to the EW signal) varying from 10\% to 2\% in the different MD regions, much smaller than the statistical uncertainty from data. This effect is taken as an uncertainty in the EW $ZZjj$ predictions. An additional uncertainty in the modelling of the QCD $ZZjj$ process is considered by comparing the predicted MD shapes from \textsc{Sherpa} to \MGNLO~2.6.1 at particle level in the case where two partons are explicitly required at LO in the ME calculation. The differences in the predicted yields in different MD regions range from $-30\%$ to $+20\%$.

\section{Observation of EW $ZZjj$}
\label{sec:ewk}

To separate the EW $ZZjj$ processes from their backgrounds, MDs based on the Gradient Boosted Decision Tree algorithm~\cite{friedman2001} are trained with simulated events using the Toolkit for Multivariate Data Analysis framework~\cite{Hocker:2007ht}. In each channel, a single MD is trained in the SR, which uses event kinematic information sensitive to the characteristics of the EW signal. The resulting MDs provide an optimal separation of the EW $ZZjj$ signal and the backgrounds, where signal-like and background-like events are featured in the high and low MD regions, respectively. In the \lllljj channel, twelve input variables are used: \mjj, \dyjj, \pt~of the leading and sub-leading jets (\ptleadj and \ptsubj), $y_{j1}\times y_{j2}$, \pt~of the $Z$ boson reconstructed from the charged-lepton pair with the mass closer to the $Z$-boson mass, rapidity of both $Z$ bosons ($y_{Z1}$ and $y_{Z2}$), \pt~and mass of the four-lepton system, \pt~of the third lepton, \pt~of the $ZZjj$ system divided by the scalar \pt~sum of $Z$ bosons and two jets ($S_{\mathrm{T}}$). Thirteen input variables are utilised in the \llvvjj channel: \mjj, \dyjj, $y_{j1}\times y_{j2}$, \ptsubj, \met, \met-significance, $S_{\mathrm{T}}$, pseudorapidity and azimuthal angle differences between two charged leptons ($\Delta\eta$, $\Delta\phi$), $\Delta R$, invariant mass of the charged-lepton pair, and \pt~of leading and sub-leading leptons. The jet-related information provides the greatest sensitivity in the \lllljj channel, while both the jet-related and the dilepton-related variables are important in the \llvvjj channel.
 
In the \lllljj channel the MD distributions in both the QCD $ZZjj$ CR and the SR are used in the statistical fit, while only the MD distribution in the SR is fitted in the \llvvjj channel. The binning of MD distributions in the SRs (Figure~\ref{fig:MVA}) is chosen to maximise the sensitivity of detecting EW $ZZjj$ events. In the \lllljj channel, the normalisation of QCD $ZZjj$ production (\muQCD) is varied simultaneously in the fit in the SR and QCD CR. The ratio of measured fiducial cross-section (with the fiducial region detailed in Section~\ref{sec:xs}) to the SM prediction for EW $ZZjj$ production (\muEW) is taken as the parameter of interest.
 
To examine the compatibility of the data and the signal-plus-background hypothesis, a test statistic is defined using the profile likelihood ratio method~\cite{Cowan:2010js}. The statistical tests are performed in both the individual \lllljj and \llvvjj channels, and in the combined channel. The experimental systematic uncertainties are considered as correlated in all the bins and regions whenever applicable. The theoretical uncertainties for $ZZjj$ production are treated as uncorrelated between the \lllljj and \llvvjj channels, due to the different fiducial volume definitions.
The QCD scale uncertainty for QCD $ZZjj$ production can be assessed in various ways in terms of correlations between different fitted regions and is conservatively treated as uncorrelated between the SR and the QCD CR in the \lllljj channel. Furthermore, the generator modelling uncertainty for QCD $ZZjj$ production is treated as uncorrelated between the low and high MD regions.
 
The results are shown in Table~\ref{tab:fit_result}. From the combined channels, the observed \muEW is $0.92\pm0.24$, while \muQCD is determined to be $0.99\pm0.22$. The statistical component accounts for 88\% of the total uncertainty in \muEW. The probability that the background can randomly fluctuate to produce a measured likelihood ratio at least as signal-like as the excess observed in the data is $1.6\times 10^{-8}$, leading to the observation of EW $ZZjj$ production. Correspondingly, with a normalised Gaussian distribution, the background-only hypothesis is rejected at 5.7~$\sigma$ (5.9~$\sigma$) from the data (expectation). The EW $ZZjj$ cross-section in the combined fiducial volume, formed by combining the respective fiducial regions in the \lllljj and \llvvjj channels, is found to be $0.70\pm0.18$~fb, calculated as \muEW multiplied by the SM prediction of $0.76 \pm 0.04$~fb.
 
\begin{table}[!htbp]
\caption{
\textbf{Significance of EW $ZZjj$ processes.} Observed \muEW and \muQCD, as well as the observed and expected significance of EW $ZZjj$ processes from the individual \lllljj and \llvvjj channels, and the combined fits.
The full set of statistical and systematic uncertainties is included.
}
\sisetup{
table-number-alignment = center,
table-align-uncertainty=true
}
\begin{center}
\begin{tabular}{
l |
S@{$\,\pm\,$}
l |
S@{$\,\pm\,$}
l |
c
}
\hline
& \multicolumn{2}{c}{$\mu_{\mathrm{EW}}$} &  \multicolumn{2}{c}{$\mu^{\ell\ell\ell\ell jj}_{\mathrm{QCD}}$}   &  Significance Obs. (Exp.) \\
\hline
\lllljj          & 0.97  & 0.27      &  0.99 & 0.22                              &  5.5 (5.6) $\sigma$     \\
\llvvjj          & 0.7  & 0.5      &  \multicolumn{2}{c|}{--}              &  1.3 (2.1) $\sigma$
\\
\hline
Combined         & 0.92 & 0.24     &  0.99 & 0.22                              &  5.7 (5.9) $\sigma$     \\
\hline
\end{tabular}
\end{center}
\label{tab:fit_result}
\end{table}
 
\begin{figure}[!htbp]
\begin{center}
 
\begin{subfigure}[b]{0.45\textwidth}
\centering
\includegraphics[width=\textwidth]{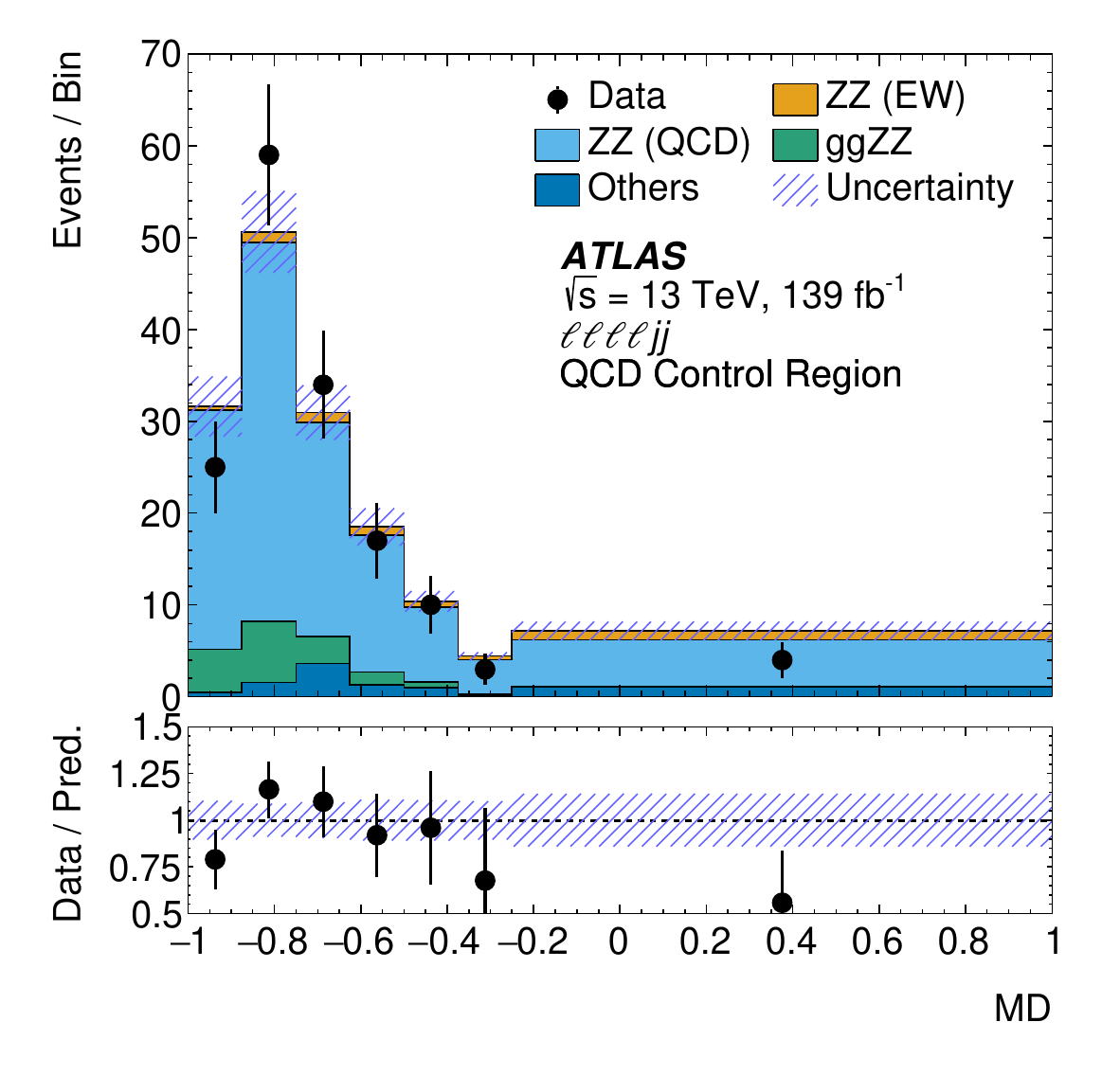}
\caption{}\label{fig:MVA_a}
\end{subfigure}
\begin{subfigure}[b]{0.45\textwidth}
\centering
\includegraphics[width=\textwidth]{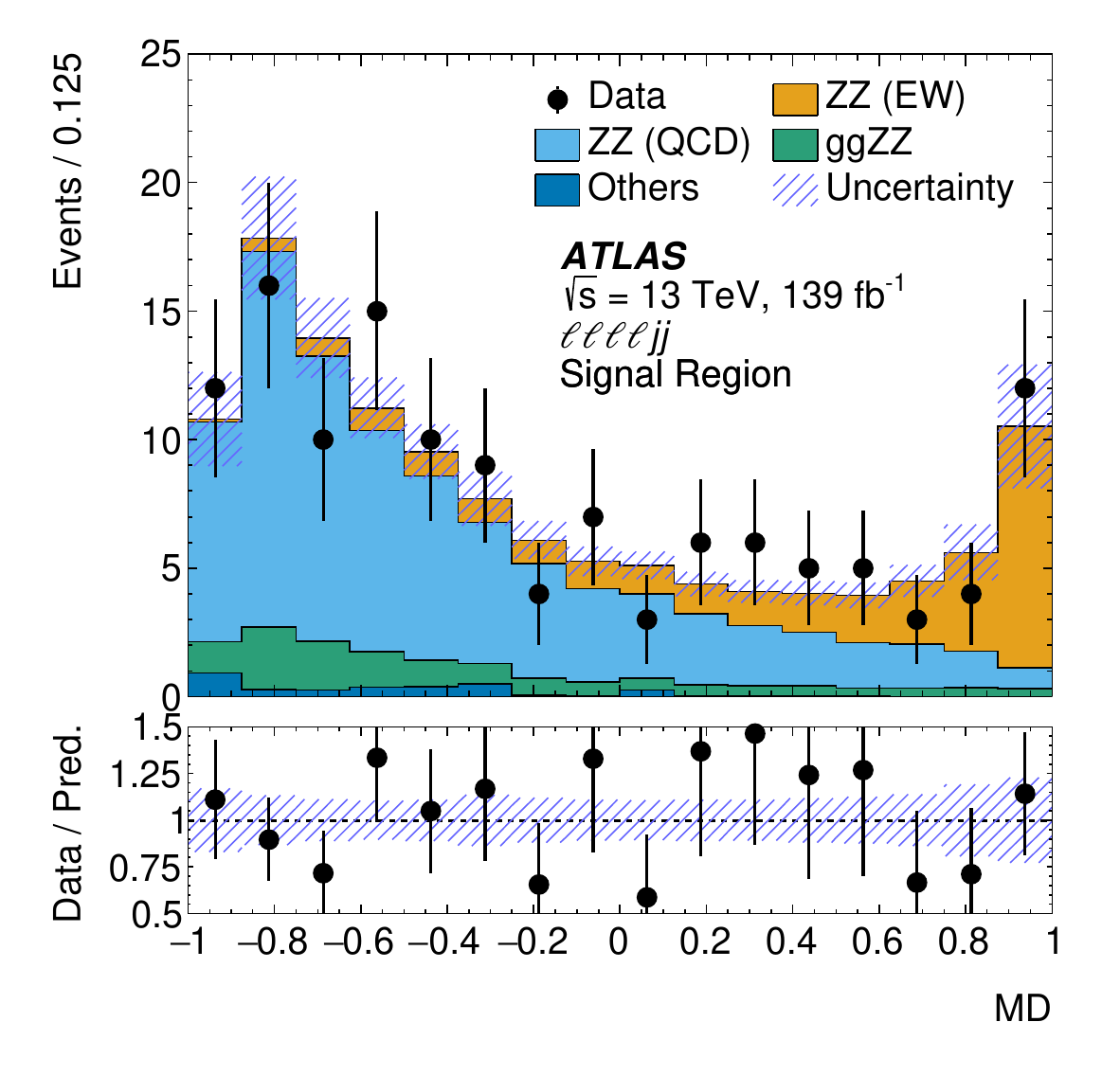}
\caption{}\label{fig:MVA_b}
\end{subfigure}
 
\begin{subfigure}[b]{0.45\textwidth}
\centering
\includegraphics[width=\textwidth]{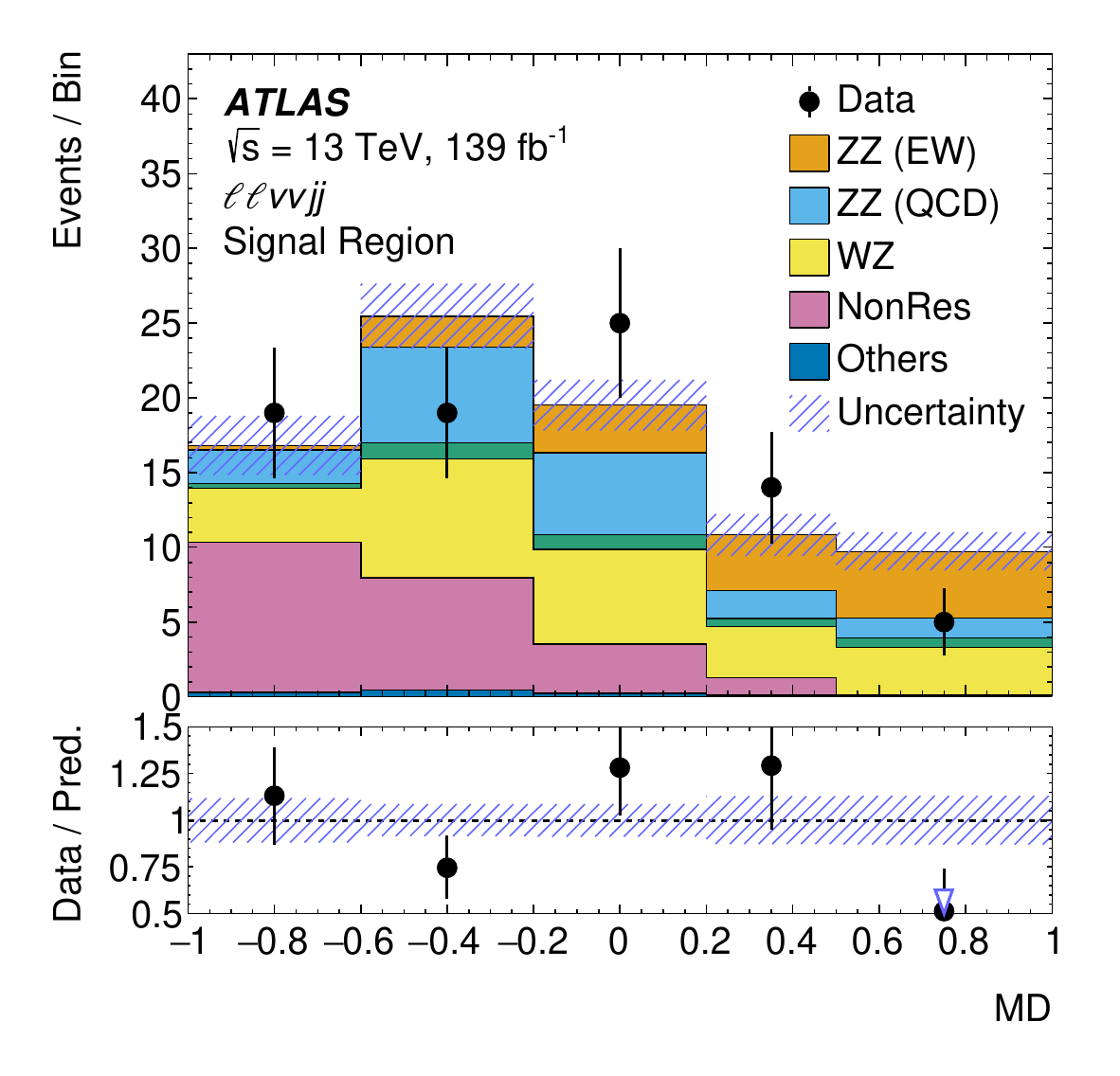}
\caption{}\label{fig:MVA_c}
\end{subfigure}
 
\end{center}
\caption{\textbf{Observed and expected multivariate discriminant distributions.}
Distributions after the statistical fit
in the (a) \lllljj QCD CR,
and in the (b) \lllljj and (c) \llvvjj signal regions.
The error bands represent standard deviations and include the experimental and theoretical systematic uncertainties,
as well as the uncertainties in \muEW and \muQCD.
The ZZ (EW), ZZ (QCD) and ggZZ represent contributions from EW, non-$gg$ QCD and $gg$ QCD $ZZjj$ processes, respectively.
The WZ represents contribution from $WZjj$ process.
All the minor backgrounds are summed together as `Others',
and the $WWjj$ and \ttbar~processes are referred to as `NonRes'.
The statistical uncertainties of the data are shown as error bars.
The open arrows represent the out-of-range markers.
The horizontal bin width is indicated on the vertical axis legend.
}
\label{fig:MVA}
\end{figure}

\section{Measurement of fiducial cross-sections}
\label{sec:xs}

In addition to the observation of the EW $ZZjj$ process, the cross-sections for the production of inclusive $ZZjj$ are also measured in the \lllljj and \llvvjj channels. This measurement, corrected for detector inefficiency and resolution without any further theoretical interpretation, provides the most model-independent results. The cross-sections are measured following the formula
$\sigma = (N_{\mathrm{data}} - N_{\mathrm{bkg}}) / (L \times C)$,
where
$N_{\mathrm{data}}$ and $N_{\mathrm{bkg}}$
refer to the number of events in data and the expected number of background events from non-$ZZjj$ processes respectively, $L$ refers to the integrated luminosity, and $C$ is the correction factor to extrapolate the QCD and EW $ZZjj$ events from detector level to the fiducial volume, calculated as the ratio of the number of $ZZjj$ events passing the detector-level event selection to the number of events selected in the fiducial volume.
 
The definitions of the fiducial volumes closely follow the detector-level selections, using `particle-level' electrons, muons, \met and jets, which are reconstructed in simulation from stable final-state particles, prior to their interactions with the detector, following the procedure described in Ref.~\cite{Aaboud:2019lgy}. In the \lllljj channel, the dilepton mass requirement is relaxed (relative to the detector-level selection) to the wider range 60--120~\GeV{} to ensure compatibility with the previous Compact Muon Solenoid publication~\cite{Sirunyan:2017fvv}. In the \llvvjj channel, both the electrons and muons are selected in the $|\eta| <$ 2.5 region to simplify the charged-lepton selections. In addition, no requirement is placed on the \met-significance due to the complexity of defining this variable at particle level; however, the particle-level \met is required to be greater than 130~\GeV{}. All the other kinematic selection requirements have the same definition as the detector-level ones.
 
The $C$-factors are found to be ($69.9 \pm 3.1$)\% in the \lllljj channel, and ($22.4 \pm 1.2$)\% in the \llvvjj channel, where the errors reflect the total uncertainties. The smaller $C$-factor in the \llvvjj channel is due to the large event migration effect in events passing the \met selection requirement at particle level that have a small \met-significance at detector level. The measured and predicted fiducial cross-sections are presented in Table~\ref{tab:xs}. Uncertainties from different sources are presented explicitly. The data statistical uncertainty dominates, while the experimental uncertainties related to jet measurements and the background estimates are the major systematic uncertainties in the \lllljj and \llvvjj channels, respectively. The measurements of $1.27 \pm 0.14$~fb for the \lllljj channel and $1.13 \pm 0.32$~fb for the \llvvjj channel are compatible with the SM predictions. The measurement precision in the \lllljj channel is better than the accuracy of the theoretical prediction.
 
\begin{table}[!htbp]
\caption{
\textbf{Measured and predicted fiducial cross-sections.} Cross-sections are presented in both the \lllljj and \llvvjj channels for the inclusive $ZZjj$ processes.
Uncertainties due to different sources are presented explicitly,
including the one from the statistical uncertainty of the data and simulated samples (stat), the one from the theoretical predictions (theo), the experimental ones due to the lepton and jet calibrations (exp), the ones from background estimates (bkg), and the one from the luminosity (lumi).
}
\begin{center}
\scalebox{0.90}{
\begin{tabular}{c | c | c}
\hline
& Measured fiducial $\sigma$ [fb] & Predicted fiducial $\sigma$ [fb] \\
\hline
\lllljj & $1.27 \pm 0.12(\mathrm{stat}) \pm 0.02(\mathrm{theo}) \pm 0.07(\mathrm{exp}) \pm 0.01(\mathrm{bkg}) \pm 0.02(\mathrm{lumi})$ & $1.26 \pm 0.04(\mathrm{stat}) \pm 0.22(\mathrm{theo})$ \\
\hline
\llvvjj & $1.13 \pm 0.28(\mathrm{stat}) \pm 0.04(\mathrm{theo}) \pm 0.06(\mathrm{exp}) \pm 0.15(\mathrm{bkg}) \pm 0.02(\mathrm{lumi})$ & $1.11 \pm 0.01(\mathrm{stat}) \pm 0.12(\mathrm{theo})$ \\
\hline
\end{tabular}}
\end{center}
\label{tab:xs}
\end{table}

\clearpage
 
\section{Outlook}
\label{sec:conclusion}

The rare electroweak production of $ZZjj$ events is observed using 139~\ifb{} of $\sqrt{s}$ = 13~\TeV{} proton--proton collision data collected with the ATLAS detector. The measurement of this rarest electroweak $VVjj$ process is an important milestone in the study of electroweak physics at the LHC. This result also marks an important step towards understanding the nature of the electroweak symmetry breaking, as it completes the observation of all major channels and confirms the consistency of the experimental results with the mechanism predicted by the Standard Model. This result marks the start of a new era in precision studies of rare processes in the electroweak sector and in searches for new phenomena that can be investigated with higher precision and in higher energy regimes with future larger datasets.

\clearpage

\FloatBarrier

\section*{Acknowledgements}


We thank CERN for the very successful operation of the LHC, as well as the
support staff from our institutions without whom ATLAS could not be
operated efficiently.
 
We acknowledge the support of
ANPCyT, Argentina;
YerPhI, Armenia;
ARC, Australia;
BMWFW and FWF, Austria;
ANAS, Azerbaijan;
SSTC, Belarus;
CNPq and FAPESP, Brazil;
NSERC, NRC and CFI, Canada;
CERN;
ANID, Chile;
CAS, MOST and NSFC, China;
Minciencias, Colombia;
MEYS CR, Czech Republic;
DNRF and DNSRC, Denmark;
IN2P3-CNRS and CEA-DRF/IRFU, France;
SRNSFG, Georgia;
BMBF, HGF and MPG, Germany;
GSRI, Greece;
RGC and Hong Kong SAR, China;
ISF and Benoziyo Center, Israel;
INFN, Italy;
MEXT and JSPS, Japan;
CNRST, Morocco;
NWO, Netherlands;
RCN, Norway;
MEiN, Poland;
FCT, Portugal;
MNE/IFA, Romania;
JINR;
MES of Russia and NRC KI, Russian Federation;
MESTD, Serbia;
MSSR, Slovakia;
ARRS and MIZ\v{S}, Slovenia;
DSI/NRF, South Africa;
MICINN, Spain;
SRC and Wallenberg Foundation, Sweden;
SERI, SNSF and Cantons of Bern and Geneva, Switzerland;
MOST, Taiwan;
TAEK, Turkey;
STFC, United Kingdom;
DOE and NSF, United States of America.
In addition, individual groups and members have received support from
BCKDF, CANARIE, Compute Canada and CRC, Canada;
COST, ERC, ERDF, Horizon 2020 and Marie Sk{\l}odowska-Curie Actions, European Union;
Investissements d'Avenir Labex, Investissements d'Avenir Idex and ANR, France;
DFG and AvH Foundation, Germany;
Herakleitos, Thales and Aristeia programmes co-financed by EU-ESF and the Greek NSRF, Greece;
BSF-NSF and GIF, Israel;
Norwegian Financial Mechanism 2014-2021, Norway;
NCN and NAWA, Poland;
La Caixa Banking Foundation, CERCA Programme Generalitat de Catalunya and PROMETEO and GenT Programmes Generalitat Valenciana, Spain;
G\"{o}ran Gustafssons Stiftelse, Sweden;
The Royal Society and Leverhulme Trust, United Kingdom.
 
The crucial computing support from all WLCG partners is acknowledged gratefully, in particular from CERN, the ATLAS Tier-1 facilities at TRIUMF (Canada), NDGF (Denmark, Norway, Sweden), CC-IN2P3 (France), KIT/GridKA (Germany), INFN-CNAF (Italy), NL-T1 (Netherlands), PIC (Spain), ASGC (Taiwan), RAL (UK) and BNL (USA), the Tier-2 facilities worldwide and large non-WLCG resource providers. Major contributors of computing resources are listed in Ref.~\cite{ATL-SOFT-PUB-2021-003}.
 

\section*{Author contributions}
All authors have contributed to the publication, being variously involved in the design and the construction of the detectors, in writing software, calibrating subsystems, operating the detectors and acquiring data, and finally analysing the processed data. The ATLAS Collaboration members discussed and approved the scientific results. The manuscript was prepared by a subgroup of authors appointed by the collaboration and subject to an internal collaboration-wide review process. All authors reviewed and approved the final version of the manuscript.
 
\section*{Competing interests}
The authors declare no competing financial interests.

\clearpage
\printbibliography
 
\clearpage
 
\begin{flushleft}
\hypersetup{urlcolor=black}
{\Large The ATLAS Collaboration}

\bigskip

\AtlasOrcid[0000-0002-6665-4934]{G.~Aad}$^\textrm{\scriptsize 102}$,    
\AtlasOrcid[0000-0002-5888-2734]{B.~Abbott}$^\textrm{\scriptsize 129}$,    
\AtlasOrcid[0000-0002-7248-3203]{D.C.~Abbott}$^\textrm{\scriptsize 103}$,    
\AtlasOrcid[0000-0002-2788-3822]{A.~Abed~Abud}$^\textrm{\scriptsize 36}$,    
\AtlasOrcid[0000-0002-1002-1652]{K.~Abeling}$^\textrm{\scriptsize 53}$,    
\AtlasOrcid[0000-0002-2987-4006]{D.K.~Abhayasinghe}$^\textrm{\scriptsize 94}$,    
\AtlasOrcid[0000-0002-8496-9294]{S.H.~Abidi}$^\textrm{\scriptsize 168}$,    
\AtlasOrcid[0000-0002-8279-9324]{O.S.~AbouZeid}$^\textrm{\scriptsize 40}$,    
\AtlasOrcid{N.L.~Abraham}$^\textrm{\scriptsize 157}$,    
\AtlasOrcid[0000-0001-5329-6640]{H.~Abramowicz}$^\textrm{\scriptsize 162}$
% ...
\bigskip
\\

% ...
$^{36}$CERN, Geneva; Switzerland.\\
% ...
$^{40}$Niels Bohr Institute, University of Copenhagen, Copenhagen; Denmark.\\
% ...
$^{53}$II. Physikalisches Institut, Georg-August-Universit\"{a}t G\"ottingen, G\"ottingen; Germany.\\
% ...
$^{94}$Department of Physics, Royal Holloway University of London, Egham; United Kingdom.\\
% ...
$^{102}$CPPM, Aix-Marseille Universit\'e, CNRS/IN2P3, Marseille; France.\\
$^{103}$Department of Physics, University of Massachusetts, Amherst MA; United States of America.\\
% ...
$^{129}$Homer L. Dodge Department of Physics and Astronomy, University of Oklahoma, Norman OK; United States of America.\\
% ...
$^{157}$Department of Physics and Astronomy, University of Sussex, Brighton; United Kingdom.\\
% ...
$^{162}$Raymond and Beverly Sackler School of Physics and Astronomy, Tel Aviv University, Tel Aviv; Israel.\\
% ...
$^{168}$Department of Physics, University of Toronto, Toronto ON; Canada.\\
% ...

\end{flushleft}


\end{document}